\documentclass[12pt]{article}
\usepackage{amsfonts,amssymb,amsmath}
\renewcommand{\text}[1]{#1}

\newcommand{\be}{\begin{equation}}
\newcommand{\ee}{\end{equation}}
\newcommand{\ben}{\begin{displaymath}}
\newcommand{\een}{\end{displaymath}}
\newcommand{\bea}{\begin{eqnarray}}
\newcommand{\eea}{\end{eqnarray}}
\newcommand{\bean}{\begin{eqnarray*}}
\newcommand{\eean}{\end{eqnarray*}}
\newcommand{\nn}{\nonumber \\}
\newcommand{\ba}{\begin{array}}
\newcommand{\ea}{\end{array}}
\newcommand{\bi}{\begin{itemize}}
\newcommand{\ei}{\end{itemize}}

\newcommand{\reef}[1]{(\ref{#1})}

\newcommand{\bbZ}{{\mathbb{Z}}}

\begin{document}

\makeatletter
\renewcommand{\theequation}{\thesection.\arabic{equation}}
\@addtoreset{equation}{section}
\makeatother

\baselineskip 18pt

\begin{titlepage}

\vfill

\begin{flushright}
DESY 08-104\\
Imperial/TP/2008/JG/01\\
\end{flushright}

\vfill

\begin{center}
   \baselineskip=16pt
  {\Large\bf $AdS$ Solutions Through Transgression}
   \vskip 1.5cm
      Aristomenis Donos$^1$, Jerome P. Gauntlett$^2$ and Nakwoo Kim$^3$\\
   \vskip .6cm
      \begin{small}
      $^1$\textit{DESY Theory Group, DESY Hamburg\\
        Notkestrasse 85, D 22603 Hamburg, Germany}
        \end{small}\\*[.6cm]
        \begin{small}
      $^2$\textit{Theoretical Physics Group, Blackett Laboratory, \\
        Imperial College, London SW7 2AZ, U.K.}
        \end{small}\\*[.6cm]
      \begin{small}
      $^2$\textit{The Institute for Mathematical Sciences, \\
        Imperial College, London SW7 2PE, U.K.}
        \end{small}\\*[.6cm]
      \begin{small}
     $^3$\textit{Department of Physics and Research Institute of
        Basic Science, \\
        Kyung Hee University, Seoul 130-701, Korea}
        \end{small}
   \end{center}

\vfill

\begin{center}
\textbf{Abstract}
\end{center}

\begin{quote}
We present new classes of explicit supersymmetric $AdS_3$ solutions of type IIB supergravity
with non-vanishing five-form flux and $AdS_2$ solutions of $D=11$ supergravity
with electric four-form flux. The former are dual to two-dimensional SCFTs with
$(0,2)$ supersymmetry and the latter to supersymmetric quantum mechanics with
two supercharges. We also investigate more general classes of $AdS_3$ solutions of type IIB supergravity and $AdS_2$
solutions of $D=11$ supergravity which in addition have non-vanishing three-form flux and
magnetic four-form flux, respectively. The construction of these more general solutions
makes essential use of the Chern-Simons or ``transgression'' terms in
the Bianchi identity or the equation of motion of the
field strengths in the supergravity theories.
We construct infinite new classes of explicit examples and for some of the type IIB solutions
determine the central charge of the dual SCFTs.  The type IIB solutions with non-vanishing three-form flux that we construct
include a two-torus, and after two T-dualities and an
S-duality, we obtain new $AdS_3$ solutions with only the NS fields being non-trivial.

\end{quote}

\vfill

\end{titlepage}
\setcounter{equation}{0}


\section{Introduction}

An interesting class of supersymmetric $AdS_3$ solutions of type IIB supergravity
with non-vanishing five-form flux and dual to $(0,2)$ SCFTs in $d=2$ were analysed in
\cite{Kim:2005ez}. Similarly, a class of $AdS_2$ solutions of $D=11$ supergravity
with electric four-form flux and dual to superconformal quantum mechanics with two
supercharges were analysed in \cite{Kim:2006qu}. It is remarkable that the geometries
of the corresponding internal seven and nine-dimensional spaces have a similar structure.
In particular, they both have a Killing vector (dual to an R-symmetry in the
corresponding SCFT)
which locally defines
a foliation, and the metrics are completely determined by a K\"ahler metric on the corresponding
six or eight-dimensional leaves. In both cases, the local K\"ahler metric satisfies the same differential equation
\be\label{m1}
\Box R-\frac{1}{2}R^{2}+{R}^{ij}{R}_{ij}=0
\ee
where $R_{ij}$ and $R$ are the Ricci tensor and Ricci scalar for the K\"ahler metric.
These $2n+1$ dimensional geometries, with $n=3,4$, were further investigated in \cite{Gauntlett:2007ts},
which also generalised them to all $n$. It was shown that the $2n+2$ dimensional cone geometries
over these spaces admit certain Killing spinors that
define an $SU(n+1)$ structure with particular intrinsic torsion
that was determined in \cite{Gauntlett:2007ts}.

This geometry has striking similarities with Sasaki-Einstein (SE) geometry.
Recall that a five-dimensional SE manifold $SE_5$ gives rise to a supersymmetric type IIB
$AdS_5\times SE_5$ solution with non-vanishing five-form flux, while a seven-dimensional
SE manifold $SE_7$ gives rise to a $AdS_4\times SE_7$ solution of $D=11$ supergravity
with electric four-form flux. All SE spaces have a Killing vector, which locally defines
a foliation, and the SE metric is completely determined by a K\"ahler-Einstein metric on the corresponding
leaves. Furthermore, the $2n+2$ dimensional cone geometries over the SE spaces are Calabi-Yau i.e. they
admit covariantly constant spinors that define
an $SU(n+1)$ structure with vanishing intrinsic torsion (i.e. the metric has
$SU(n+1)$ holonomy).

The $AdS_5\times SE_5$ and $AdS_4\times SE_7$ solutions are the near
horizon limits of more general supergravity solutions that describe
$D3$-branes and $M2$-branes sitting at the apex of the Calabi-Yau
three and four-fold cones, respectively. In these more general
solutions, only the five-form flux and electric four-form flux are
non-trivial, and the solutions are determined by a harmonic function
on the Calabi-Yau space. An interesting further generalisation for
the type IIB case, is to consider any Calabi-Yau three fold and to
switch on imaginary self-dual harmonic three form flux. One finds
that this solution preserves the same amount of supersymmetry.
Furthermore the Bianchi identity for the five-form, modified by
Chern-Simons or ``transgression'' terms,
\be\label{iibt} dF_5=\frac{i}{2}G\wedge G^* \ee where $G$ is a
complex three-form which contains the NS-NS and R-R three-forms,
implies that the solutions are determined by a function that
satisfies a Laplace equation with a source term. Similarly, for
$D=11$ supergravity one can consider an arbitrary Calabi-Yau
four-fold and switch on a harmonic self-dual four-form. Now it is
the equation of motion for the three-form potential with its
transgression terms, \be\label{d11t} d*_{11}G_4+\frac{1}{2}G_4\wedge
G_4=0 \ee which is playing a key role in the solution. Switching on
the additional fluxes in these type IIB and $D=11$ solutions necessarily breaks
the conformal symmetry. A prominent example of such solutions is
the Klebanov-Strassler solution of type IIB \cite{Klebanov:2000hb}
(see also \cite{Klebanov:2000nc,Grana:2000jj}), which is constructed
using the deformed conifold metric. A more general analysis of these
kinds of solutions can be found in \cite{Cvetic:2000mh}.

One of the main aims of this paper is to show that we can similarly generalise the classes of
type IIB solutions considered in \cite{Kim:2005ez} and the $D=11$ solutions considered in \cite{Kim:2006qu}
to include three-form flux and magnetic four-form flux, respectively.
The central idea is to switch on such fluxes on the six and eight dimensional K\"ahler spaces, respectively.
We will show that this can be done in a way that maintains the $AdS_3$ and $AdS_2$ factors,
and hence the dual conformal symmetry (in contrast to the examples discussed above),
and also preserves the same amount of supersymmetry.
We find that the solutions are still, locally, specified by a K\"ahler
metric but \reef{m1} is modified by a term involving the new activated fluxes.
We will also construct rich new classes of explicit solutions by following a similar analysis to
that of \cite{Gauntlett:2006ns}.

The plan of the rest of the paper is as follows. We will summarise the
general classes of $AdS_3$ solutions of type IIB and $AdS_2$ solutions of $D=11$ in section 2. We have
left some details of the derivations, which are very similar to those in \cite{Kim:2005ez}
and \cite{Kim:2006qu}, to appendix A. We will also briefly interrupt the main narrative to explain how the solutions can be analytically
continued so that the $AdS$ factors are replaced by spheres. This gives rise to new general classes
of 1/8 BPS bubble solutions generalising those discussed in \cite{Gauntlett:2006ns}
(1/2 BPS bubble solutions were first analysed in \cite{Lin:2004nb}, and
other studies of general classes of bubble solutions
preserving various amounts of
supersymmetry in type IIB and $D=11$ supergravity have appeared in
\cite{Yamaguchi:2006te}-\cite{D'Hoker:2008wc}).

In section 3 we will construct explicit $AdS$ solutions by taking the six and eight dimensional K\"ahler metrics to be products of two-dimensional K\"ahler-Einstein (KE) spaces. For type IIB we will first analyse the global properties of
the local solutions with vanishing three-form flux that were found in \cite{Gauntlett:2006ns} and calculate the
central charge of the dual CFTs. These $AdS_3$ solutions are labelled by a rational number $s/t\in[-1/2,0)$ and an integer $N$ fixing
the five-form flux. The topology of the internal seven-manifold is a certain $U(1)$ bundle over a product of two two-spheres
and a Riemann surface with genus greater than one.
We then consider solutions with non-zero three-form flux by taking one of the K\"ahler-Einstein factors to be a two-torus. We find that the two other KE
spaces must be spheres. After two T-dualities
we find that the solutions turn out to be the well known $AdS_3\times S^3\times S^3\times S^1$ solutions of type
IIB supergravity (see \cite{Cowdall:1998bu,Boonstra:1998yu,deBoer:1999rh,Gukov:2004ym}).
We conclude section 3 with a similar
construction of explicit $AdS_2$ solutions of $D=11$ with non-vanishing magnetic four-form flux.

In sections 4 and 5 we will present a different construction of local six and eight dimensional K\"ahler metrics,
using fibrations over KE spaces, generalising the constructions in \cite{Gauntlett:2006ns} (see also \cite{Gauntlett:2006af}). We have recorded
some details in appendices C and D, respectively.
For type IIB we will consider the product of $T^2$ with a two-dimensional fibration over an $S^2$. This leads to infinite new explicit examples of
$AdS_3$ solutions of type IIB supergravity with the internal seven dimensional space having topology $S^3\times S^2\times T^2$ and the metric
labelled by a pair of positive relatively prime integers $p,q$. When the type IIB
three-flux is vanishing we show that demanding that the five-form is properly quantised implies that as solutions of type IIB string theory
they depend on two more integers $M,N$ which fix the five-form flux and the size of the $T^2$. For these solutions we calculate the
central charge of the dual CFTs\footnote{The corresponding analysis for the case when the three-form flux is non-vanishing will be determined in \cite{ajj}.}.
We also show that after two T-dualities the solutions are mapped to type IIB $AdS_3$ solutions with non-vanishing dilaton and RR three-form: after a further S-duality
only NS fields are non-zero.

Section 5 carries out similar constructions of local eight dimensional K\"ahler metrics which are the product of $T^2$ with
a two-dimensional fibration over a four dimensional KE space with positive curvature. This gives rise to infinite classes
of $AdS_2$ solutions with non-vanishing magnetic four-form flux.
Section 6 briefly concludes.

\section{$AdS$ solutions through transgression}

We first consider a general class of supersymmetric $AdS_3$ solutions of type IIB supergravity that are dual to $(0,2)$ SCFTs in $d=2$.
The metric and the self-dual five-form take the form
\bea\label{iiboa}
ds^{2} & =&e^{2A}\left[ds^{2}\left(AdS_{3}\right)+ds^{2}(Y_7)\right]\nn
F_{5} & =&\left(1+\ast_{10}\right)\text{Vol}({AdS_{3}})\wedge F_{2}
\eea where $F_2$ is a two-form on $Y_7$.
The dilaton and axion are constant and for simplicity we set them to  zero.
We also demand that the complex three-form flux, $G$, which contains the
NS-NS and R-R three-form field strengths, is a three-form on $Y_7$.

As we show in appendix A, by following the analysis of
\cite{Kim:2005ez}, demanding that this is a supersymmetric solution
to the equations of motion,
preserving supersymmetry as described in the appendix,
leads to the following local description.
The metric can be written \be\label{bit} ds^{2}(Y_7)=
\frac{1}{4}\left(dz+P\right)^{2}+e^{-4A}ds^2_6 \ee where
$\partial_z$ is a Killing vector, $ds^2_6$ is a K\"ahler metric
and $dP$ is the Ricci form for $ds^2_6$. The warp factor is given by
\be e^{-4A}=\frac{1}{8}R \ee where $R$ is the Ricci scalar for
$ds^2_6$ and we thus need to demand that $R>0$. The two-form $F_2$ appearing in the five-form
can be written
 \be\label{iib4}
F_{2}=
{2}J-\frac{1}{2}d\left[e^{4A}\left(dz+P\right)\right] \ee
where $J$ is the K\"ahler form for $ds^2_6$.

So far, this is exactly the same as when the three-form flux vanishes \cite{Kim:2005ez}.
However, further analysis shows that we can switch on the three-form $G$, provided
that $G$ is a closed, $(1,2)$ and primitive three-form on the
K\"ahler space. In particular $G$ must be imaginary self-dual,
$*_6G=iG$, and harmonic. Furthermore, the Bianchi identity for the
five-form with its transgression terms \reef{iibt}, implies that the
K\"ahler metric $ds^2_6$ must satisfy
\be\label{miib} \Box
R-\frac{1}{2}R^{2}+{R}^{ij}{R}_{ij}+\frac{2}{3}G^{ijk}G_{ijk}^{\ast}=0
\ee
which is the key equation generalising \reef{m1}.

We now consider a general class of supersymmetric $AdS_2$ solutions of $D=11$ supergravity that are dual
to superconformal quantum mechanics with two supercharges.
The metric and the four-form are given by
\bea\label{pop}
ds^{2} & =&e^{2A}\left[ds^{2}\left(AdS_{2}\right)+ds^{2}(Y_9)\right]\nn
G_{4} & =&\mathrm{Vol}({AdS_{2}})\wedge F_{2}+{F}_{4}
\eea
where $F_2$ is a two-form on $Y_9$ and $F_4$ is a four-form on $Y_9$.
This generalises the class of solutions studied in \cite{Kim:2006qu} which had $F_4=0$ i.e. purely electric fluxes.

As we show in appendix A, now following the analysis of \cite{Kim:2006qu}, demanding that this is a supersymmetric solution to the equations
of motion, preserving supersymmetry as described in the appendix,
leads to the following local description. The metric can be written
\be\label{bob}
ds^{2}(Y_9)=  \left(dz+P\right)^{2}+e^{-3A}ds^2_8
\ee
where $\partial_z$ is a Killing vector, $ds^2_8$ is a K\"ahler metric and $dP$ is the Ricci form for $ds^2_8$.
The warp factor is given by
\be
e^{-3A}=\frac{1}{2}R
\ee
where $R$ is the Ricci scalar for $ds^2_8$ and so we demand $R>0$.
The two-form $F_2$ appearing in the four-form can be written
\be\label{pop4}
F_{2}=-J+ d\left[e^{3A}\left(dz+P\right)\right]
\ee
where $J$ is the K\"ahler form for $ds^2_8$.
This is exactly as in the case of purely electric four-form flux \cite{Kim:2006qu}. We now find that we can switch
on $F_4$ provided that it is a closed, $(2,2)$ and primitive four-form on the K\"ahler space.
In particular $F_4$ must be self-dual and harmonic.
Furthermore, the equation of motion for the four-form with its transgression terms \reef{d11t} implies that
the K\"ahler metric $ds^2_8$ must now satisfy
\be\label{md11}
\Box R-\frac{1}{2}R^{2}+{R}^{ij}{R}_{ij}+\frac{1}{4!} F_4^{ijkl} {F}_{4ijkl}=0.
\ee

In the special case that the eight-dimensional K\"ahler metric $ds^2_8$
contains a $T^2$ factor, we can dimensionally reduce the $D=11$ solution
on one leg of the $T^2$ and then T-dualise on the other leg, to obtain a type IIB solution.
In the case that $F_4=0$, it was shown in \cite{Gauntlett:2006ns} that the resulting
type IIB solution is in fact the $AdS_3$ solution with vanishing three-form flux. There is a simple generalisation
to non-vanishing $F_4$. Decompose the eight-dimensional K\"ahler form as
\be
J_{8}=J_{6}+du^1\wedge du^2
\ee
where $u^1,u^2$ are coordinates on the $T^2$. Suppose we can write the $\left(2,2\right)$ four form as
\be
{F}_{4}={i}G\wedge\left(du^1+i du^2\right)-{i}G^{\ast}\wedge\left(du^1-i du^2\right)
\ee
where $G$ is a closed primitive $(1,2)$ form in six-dimensions (i.e. we are asuuming that
there is no term involving the volume form of the two torus, $du^1\wedge du^2$). If we dimensionally reduce on the $u^2$ direction and
then T-dualise on the $u^1$ direction we find that the $D=11$ $AdS_2$ solution is transformed into
the type IIB $AdS_3$ solution.

\subsection{Bubble solutions}

In subsequent sections we will find explicit examples of the $AdS_3$ and $AdS_2$ solutions just described.
Before doing that we pause to briefly comment on how the above classes of solutions can be analytically continued
so that the $AdS$ factors are replaced with spheres. These ``bubble'' solutions preserve 1/8 of the supersymmetry
and generalise those discussed in \cite{Gauntlett:2006ns}.

For the type IIB case, the metric is given by
\be
ds^{2}  =e^{2A}\left[-\frac{1}{4}\left(dt+P\right)^{2}
+ds^{2}\left(S^3\right)+e^{-4A}ds^2_6\right]\ee
where $\partial_t$ is a Killing vector, $ds^2_6$ is again a K\"ahler metric and
$dP$ is the Ricci form for $ds^2_6$. The warp factor is given by
\be e^{-4A}=-\frac{1}{8}R \ee where $R$ is the Ricci scalar for
$ds^2_6$ and so now we want $R<0$. The five-form flux
is given by
\be
F_{5}  =\left(1+\ast_{10}\right)\text{Vol}(S^3)\wedge F_{2}
\ee
where
\be F_{2}=
{2}J+\frac{1}{2}d\left[e^{4A}\left(dt+P\right)\right] \ee
and $J$ is the K\"ahler form for $ds^2_6$.
The three-form $G$ is again a closed, $(1,2)$ and primitive three-form on the
K\"ahler space. Finally the master equation reads
\be
\Box R-\frac{1}{2}R^{2}+{R}^{ij}{R}_{ij}-\frac{2}{3}G^{ijk}G_{ijk}^{\ast}=0.
\ee

For the $D=11$ case, the metric is given by
\be
ds^{2}=  e^{2A}\left[-\left(dt+P\right)^{2}+ds^2(S^2)+e^{-3A}ds^2_8\right]
\ee
where $\partial_t$ is a Killing vector, $ds^2_8$ is a K\"ahler metric and $dP$ is the Ricci form for $ds^2_8$. The warp
factor is given by
\be
e^{-3A}=-\frac{1}{2}R
\ee
where $R$ is the Ricci scalar for $ds^2_8$ and we demand $R<0$.
The four-form flux is given by
\be
G_4=Vol(S^2)\wedge F_2 +F_4
\ee
where
\be
F_{2}=-J- d\left[e^{3A}\left(dt+P\right)\right]
\ee
and $J$ is the K\"ahler form for $ds^2_8$.
$F_4$ is again a closed, $(2,2)$ and primitive four-form on the K\"ahler space.
Finally, the master equation is now
\be
\Box R-\frac{1}{2}R^{2}+{R}^{ij}{R}_{ij}-\frac{1}{4!} F_4^{ijkl} {F}_{4ijkl}=0.
\ee

\section{Product of KE spaces}

In this section we will explore solutions for which the K\"ahler metrics $ds^2_6$ and $ds^2_8$
appearing in \reef{bit} and \reef{bob}, respectively,
are simply the product of a set of two-dimensional K\"ahler--Einstein
metrics
\be\label{prodans}
   ds^2_{2n} =\sum_{i=1}^{n} ds^2(KE^{(i)}_2)
\ee
where $ds^2(KE^{(i)}_2)$ is a two-dimensional K\"ahler-Einstein metric,
i.e. locally proportional to the standard metric on $S^2$, $T^2$ or
$H^2$. For the latter case, we can also take a quotient $H^2/\Gamma$ to get
a Riemann surface with genus greater than one.
The metric $ds^2_{2n}$ is normalised so that the Ricci form is given by
\be\label{prodans2}
{\cal R}=\sum_{i=1}^{n}{\cal R}_i=\sum_{i=1}^{n}l_i J_i
\ee
where ${\cal R}_i$ and $J_i$ are the Ricci and K\"ahler forms of the $ds^2(KE_2^{(i)})$
metrics, respectively, and $l_i$ is zero, positive or negative depending on whether
the metric is locally that on $T^2$, $S^2$ or $H^2$, respectively.
We also have $P=\sum_i P_i$ with $dP_i= {\cal R}_i$ and the Ricci scalar is
$R=2\sum_{i=1}^{n}l_i$.
Note that in the special case that two of the $l_i$ are equal, say
$l_1=l_2$, the analysis can be simply extended to cover the case when the product
$KE_2^{(1)}\times KE_2^{(2)}$ is replaced with a more general
four-dimensional K\"ahler-Einstein manifold, $KE_4$. Similar
generalisations are possible if more of the $l_i$ are equal.
Finally, it will be useful to recall that if the $i$th KE space,
$\Sigma_{g_i}$, is a Riemann surface of genus $g_i$, then
\be
\frac{1}{2\pi}\int_{\Sigma_{g_i}} {\cal R}_i=2(1-g_i)\ .
\ee

\subsection{Type IIB}
For this case, the metric $ds^2(Y_7)$ appearing in \reef{iiboa}
is given by
\be
\frac{1}{L^2}ds^{2}(Y_7)=\frac{1}{4}\left(dz+P\right)^{2}+e^{-4A}\left[
\sum_{i=1}^{3} ds^2(KE^{(i)}_2)
 \right] \ee
where we have introduced an overall length scale $L$, and the warp factor is given by
\be\label{warpy}
e^{-4A}=\frac{1}{4}(l_1+l_2+l_3)\ . \ee
Writing the five-form flux as
\be\label{fff1} F_5=AdS_3\wedge F_2+\omega_5 \ee we have
\bea\label{fiveformflux2} \frac{1}{L^4}F_2&=&
\frac{2}{l_1+l_2+l_3}\left[(l_2+l_3)J_1+(l_1+l_3)J_2+(l_1+l_2)J_3\right]\nn
\frac{1}{L^4}\omega_5&=&\frac{1}{4}\left[
(l_1+l_2)J_1\wedge J_2+(l_1+l_3)J_1\wedge J_3+(l_2+l_3)J_2\wedge J_3\right](dz+P)\ . \eea

\subsubsection{$G=0$}
We first consider the local solutions with zero three-form flux, $G=0$,
that were presented in section 6.1 of \cite{Gauntlett:2006ns}.
We will show that there are an infinite number of globally defined solutions
with appropriately quantised five-form flux and we will calculate the
central charges of the dual $d=2$ $(0,2)$ SCFTs.

It was shown in \cite{Gauntlett:2006ns} that the master equation
\reef{m1} is solved if $(l_1,l_2,l_3)$=$(l_1,-\frac{l_1}{1+l_1},1)$ with $l_1\in[-1/2,0]$.
When $l_1=0$ we obtain the well known $AdS_3\times S^3\times T^4$ solution.
We therefore restrict to $l_1\in [-1/2,0)$ so that the six-dimensional
K\"ahler manifold is $\Sigma_g\times S^2_1\times S^2_2$, where
$\Sigma_g$ is a Riemann surface with genus $g>1$.

We now examine the conditions required for $Y_7$ to be a well defined $U(1)$ fibration over
$\Sigma_g\times S^2_1\times S^2_2$. If we let the period of the coordinate $z$ be $2\pi l$ then we
require that $l^{-1}P$ be a bona-fide $U(1)$ connection. This is guaranteed if the integral of
$l^{-1}dP/(2\pi)$ over a basis of two cycles on $\Sigma_g\times S^2_1\times S^2_2$ are all integers.
Taking the obvious basis, we conclude that we should take $z$ to have period
$4\pi$ and then the periods are $(1-g,1,1)$.

We now turn to the five-form. We first observe that this is a
globally defined five-form on $Y_7$. To ensure that we have a good
solution of type IIB string theory, we demand that the five-form
flux is properly quantised:
\bea N(D)=\frac{1}{(2\pi
l_s)^4g_s}\int_DF_5\in \bbZ \eea for any five-cycle $D\in
H_5(Y_7,\bbZ)$. A basis for the free part of $H_5(Y_7,\bbZ)$ is
obtained by taking the $U(1)$ fibration over a basis of four-cycles
on the base $\Sigma_g\times S^2\times S^2$. Let $D_1$, $D_2$ and
$D_3$ denote the five cycles arising from the four-cycles
$\Sigma_g\times S^2_1$, $\Sigma_g\times S^2_2$ and $S^2_1\times
S^2_2$, respectively. Since the $U(1)$ fibration is non-trivial,
these five-cycles are not independent in homology and we have
$[D_1]+[D_2]+(1-g)[D_3]=0$. Calculating the $N(D_i)$ we then deduce that
for them to be all integers, $l_1$ must be rational, $l_1=s/t$ and
\be \frac{L^4}{\pi g_s l_s^4}=\frac{s}{h}N \ee where
$h=hcf(t,(g-1))$. Indeed, we then find that \bea
N(D_1)&=&-\frac{s(1-g)}{h}N\nn N(D_2)&=&\frac{(s+t)(1-g)}{h}N\nn
N(D_3)&=&-\frac{t}{h}N\ . \eea Clearly we have
$N(D_1)+N(D_2)+(1-g)N(D_3)=0$ which corresponds to the relation
amongst the five-cycles mentioned above.

We have thus established that there is an infinite class of solutions
labelled by rational $l_1=s/t\in [-1/2,0)$, each of which gives rise
to a $d=2$ $(0,2)$ SCFT. The central charge of the SCFTs is given by
\be\label{bh}
c=\frac{3R_{AdS_3}}{2G_{(3)}}
\ee
where $G_{(3)}$ is the three-dimensional Newton's constant and
$R_{AdS_3}$ is radius of the $AdS_3$ space.
In our conventions the type IIB supergravity Lagrangian
has the form
\be\label{bhconv}
\frac{1}{(2\pi)^7g_s^2l_s^8}{\sqrt{-det g}}R+\dots
\ee
and we calculate that
\be
c=6\frac{(g-1)(s^2+st+t^2)}{h^2}N^2\ .
\ee
Note that for the special case of $s=1,t=-2$ we have $(l_1,l_2,l_3)=(-1/2,1,1)$:
this is a case whose central charge was already calculated in
\cite{Gauntlett:2006qw} (substitute $M=8$, $m=2$ into equation (6.14) of that reference).

\subsubsection{$G\ne 0$}

We now turn to the construction of solutions with non-vanishing three-form flux.
In order to find a suitable three-form flux $G$ we  will demand that the product of the KE spaces includes a $T^2$ factor, $l_3=0$.
We then take the three-form to be given by
\be
\frac{1}{L^2}G=d\bar u\wedge [m_1J_1+m_2J_2]
\ee
where $u$ is a complex coordinate on the $T^2$ and $m_1, m_2$ are constant.
This is closed and is also a $(1,2)$ form on the K\"ahler space.
In order that it is primitive we must set $m_1=-m_2$. Without loss of
generality we take $m_1>0$. It just remains to solve the
master equation \reef{miib} which gives
\be
l_1l_2=4m_1^2\ .
\ee
Recalling the expression for the warp factor, \reef{warpy} (with $l_3=0$),
which must be positive, we deduce that
$l_i>0$ and in particular our six-dimensional K\"ahler space must be
$S^2_1\times S^2_2\times T^2$. After a possible rescaling we can take
$l_2=1$. The five-form flux is given by \reef{fff1} and \reef{fiveformflux2} with $l_3=0$.

To analyse this solution further, it is convenient to perform successive T-dualities on the two legs of the
$T^2$ (which we take to be square). Using the formulae in appendix B, we are led to the following type IIB
solution\footnote{To obtain the solution in this form, we rescaled the $u^1$ coordinate, $u^1\to u^1 (m_1/l_2)$,
we set the dilaton to zero by shifting the dilaton and rescaling $F_3$, and we also absorbed the warp factor into $L^2$.}
\bea
\frac{1}{L^2}ds^2&=&ds^2(AdS_3)+\frac{a+1}{4a}ds^2(S^2_1)+\frac{a+1}{4}ds^2(S^2_2)\nn
&&+\frac{1}{4}(dz+P_1+P_2)^2+\frac{a}{4}(d {u^1}-\frac{1}{a}P_1
+P_2)^2+(du^2)^2\nn
\frac{1}{L^2}F_3&=&2Vol(AdS_3)+\frac{1}{4}({\cal R}_1+{\cal
R}_2)(dz+P_1+P_2) -\frac{a}{4}(\frac{1}{a}{\cal R}_1-{\cal
R}_2)(du^1-\frac{1}{a}P_1+P_2)\nn e^{2\phi}&=&1\ . \eea Note that here
(unlike above) the metrics on the two-spheres have unit radius and
$a=l_1/l_2$.
Introducing the coordinates $\psi_1=(a/(1+a))(z-y)$ and $\psi_2=(1/(1+a))(z+ay)$ and then
completing the squares using the $\psi_i$ we are led to
\bea
\frac{1}{L^2}ds^2&=&ds^2(AdS_3)+\frac{a+1}{a}ds^2(S^3_1)+(a+1)ds^2(S^3_2)+(du^2)^2\nn
\frac{1}{L^2}F_3&=&2Vol(AdS_3)+\frac{2(a+1)}{a}Vol(S^3_1)
+2(a+1)Vol(S^3_2)
\eea
where $ds^2(S^3_i)$ are the round metrics on unit radius three spheres.
This is the well known $AdS_3\times S^3\times S^3\times S^1$ solution of type
IIB supergravity (see \cite{Cowdall:1998bu,Boonstra:1998yu,deBoer:1999rh,Gukov:2004ym}).
Note that this solution is dual to a $d=2$ SCFT with $(4,4)$
supersymmetry: when we T-dualise back the configuration with $G\ne 0$ we will
possibly break some of
the supersymmetry: our construction guarantees that there is at least $(0,2)$ supersymmetry, but we haven't
checked if more supersymmetry is preserved.

\subsection{$D=11$}
We briefly consider similar constructions of $AdS_2$ solutions of $D=11$ supergravity.
The metric $ds^2(Y_9)$ appearing in \reef{pop}
is given by
\be
ds^{2}(Y_9)=\left(dz+P\right)^{2}+e^{-3A}\sum_{i=1}^{4} ds^2(KE^{(i)}_2)
\ee
and the warp factor is given by
\be
e^{-3A}=\sum_{i=1}^4 l_i\ . \ee
The four form flux is
\be\label{4ff1} G_4=Vol(AdS_2)\wedge F_2+F_4 \ee with
\bea\label{fourformflux2} F_2&=&
\frac{2}{\sum_{i=1}^4l_i}\left[(l_1+l_2+l_3)J_1+(l_1+l_3+l_4)J_2+(l_1+l_2+l_4)J_3+(l_1+l_2+l_3)J_4\right]\nn
{F}_{4}&=&\sum_{i,j}m_{ij}\, J^{i}\wedge J^{j}
\eea
where the entries of the symmetric matrix $m$ are constants and the diagonal entries are zero.
Clearly $F_4$ is a $(2,2)$ form. Demanding that it is primitive implies that\begin{equation}
m_{12}=m_{34},\qquad m_{13}=m_{24},\qquad m_{14}=m_{23}\ ,\label{eq:m_constrains}\end{equation}
and hence $F$ is self dual, and
\be
m_{12}+m_{13}+m_{14}=0\ .
\ee
Finally, the master equation \reef{md11} now implies that
\be
l_1l_2+l_1l_3+l_1l_4+l_2l_3+l_2l_4+l_3l_4=2[(m_{12})^2+(m_{13})^2+(m_{14})^2]\ .
\ee

In the special case that one has a $T^2$ factor, say $l_4=0$, one might wonder if one can get a
type IIB $AdS_3$ solution after dimensional reduction and T-duality.
Following the discussion at the end of section 2, in order to get
an $AdS_3$ factor one needs that $m_{i4}=0$ for all $i$.
This implies all the $m_{ij}=0$ and one returns to the cases analysed
in \cite{Gauntlett:2006ns}.

\section{Fibration Constructions using $KE$ spaces: type IIB solutions}

In this section we will construct new $AdS_3$ solutions of type IIB supergravity both with $G=0$ and $G\ne 0$.
For both cases we will take the local six-dimensional dimensional K\"ahler metric, $ds^2_6$, to be
the product of $T^2$ with a four dimensional local K\"ahler metric which is constructed using the line bundle over
a two dimensional K\"ahler Einstein space, which we take to be an $S^2$. The construction of such K\"ahler spaces
is very similar to the construction in section 3 of \cite{Gauntlett:2006ns} which in turn was inspired by \cite{Page:1985bq}.
Using this construction we take $G$ to be the wedge product of a $(0,1)$ form on the $T^2$ with a $(1,1)$ form on the four-dimensional
K\"ahler space. We have presented a few details of the derivation of these solutions in appendix C.

The metric of type IIB supergravity is given by
\be
\frac{1}{L^2}ds^2=\frac{\beta}{y^{1/2}}[ds^2(AdS_3)+ds^2(Y_7)]
\ee
where $L$ is an arbitrary length scale,
\bea\label{mettues}
ds^2(Y_7)&=&\frac{\beta^2-1+2y-Q^2y^2}{4\beta^2}Dz^2+\frac{U(y)}{4(\beta^2-1+2y-Q^2y^2)}D\psi^2
+\frac{dy^2}{4\beta^2y^2 U(y)}\nn
&&+\frac{1}{\beta^2}ds^2(S^2)+\frac{y}{\beta^2}ds^2(T^2)
\eea
with $D\psi=d\psi+2V$, $dV=2J_{S^2}$  and the round metric on $S^2$, $ds^2(S^2)$,  is normalised so that
${\cal R}_{S^2}=4J_{S^2}$.
We also have
\be
Dz=dz-g(y)D\psi
\ee
with
\be
g(y)=\frac{y(1-Q^2y)}{\beta^2-1+2y-Q^2y^2}
\ee
and
\bea
U(y)=1-\frac{1}{\beta^2}(1-y)^2-Q^2y^2
\eea
where $\beta, Q$ are positive constants.

The self-dual five-form can be written \bea F_5=AdS_3\wedge
F_2+\omega_5 \eea with
 \bea
\frac{1}{L^4}F_2=\frac{\beta^2(1-Q^2y)}{2y(\beta^2-1+2y-Q^2y^2)}dy\wedge D\psi
+\frac{\beta^2}{2y^2}dy\wedge Dz+2J_{S^2}+2Vol(T^2) \eea and
\bea
\frac{1}{L^4}\omega_5&=&
-\frac{y(1-Q^2y)}{\beta^2}Vol(T^2)\wedge J_{S^2}\wedge  Dz +\frac{U(y)}{(\beta^2-1+2y-Q^2y^2)}Vol(T^2)\wedge J_{S^2}\wedge D\psi\nn
&&
-\frac{1}{4 \beta^2y^2}dy\wedge D\psi\wedge J_{S^2}\wedge Dz-\frac{1}{4\beta^2} Vol(T^2)\wedge dy\wedge D\psi\wedge Dz\ . \eea
If we introduce a complex coordinate $u=u^1+iu^2$ on the
$T^2$ with $ds^2(T^2)=dud\bar u$, we can write the three-form
flux as \bea \frac{1}{L^2}G=\frac{Q}{\beta}{d\bar
u}\wedge \left[\frac{(1-g)}{2}dy\wedge D\psi-\frac{1}{2}dy\wedge Dz+2yJ_{S^2}\right]\ . \eea

We now investigate how to restrict the parameters $(\beta,Q)$ and choose suitable ranges of the coordinates
so that these local solutions can be extended to provide good globally defined solutions.
In section 4.1, for $G=0$, we show that there are an infinite number of solutions of type IIB string theory,
labelled by a pair of positive relatively prime integers, $p,q$, and two integers $M,N$
where $Y_7$ has topology $S^3\times S^2\times T^2$. The five-form flux is
properly quantised an we also calculate the central charge of the corresponding dual CFTs.
In section 4.2, for $G\ne 0$, we show that there is a similar infinite class of $AdS_3$ solutions of type IIB supergravity,
but the analysis of the flux quantisation will be studied in \cite{ajj}.
In section 4.3 we show that after two T-dualites and an S-duality all of these solutions get transformed
into type IIB solutions with only NS fields being non-trivial.

\subsection{Type IIB solutions with $G=0$}
Setting $Q=0$ so that
\be
U(y)=1-\frac{1}{\beta^2}(1-y)^2
\ee
we choose
\be
y_1\le y\le y_2
\ee
where $y_i$ are two positive distinct roots of $U$.
The roots of $U$ are given by
\be
y_1=1-\beta,\qquad y_2=1+\beta
\ee
and we therefore choose $0<\beta<1$.

We want to argue, after suitable further restrictions, that $Y_7=M_5\times T^2$ is
the product of a two-torus with a five manifold $M_5$, parametrised by
$z,y,\psi$ and the round $S^2$.
More precisely the manifold $M_5$ will be a good
circle fibration, with the fibre coordinate labelled by $z$,
over a four-dimensional base manifold, $B_4$,
parametrised by $y,\psi$ and the round $S^2$. The analysis is very similar to that for the five-dimensional Sasaki-Einstein
metrics of \cite{Gauntlett:2004yd} (for further dicussion see \cite{Martelli:2004wu}).

We first observe that if we choose the period of $\psi$ to be $2\pi$, then $y,\psi$ parametrise a smooth two-sphere
(in particular, one can check that there are no conical singularities at the poles $y=y_1$ and $y=y_2$) and that
$B_4$ is a smooth manifold which is an $S^2$ bundle over the round $S^2$. In fact, topologically, $B_4=S^2\times S^2$.
To construct $M_5$ as a circle bundle over $B_4$, we let $z$
be periodic with period $2\pi l$. We next observe that the norm of the
Killing vector $\partial_z$ is non-vanishing and so the size of the
$S^1$ fibre doesn't degenerate. If we write $Dz=dz-A$, we require
that $l^{-1}A$ is a connection on a bona fide $U(1)$ fibration.
This is guaranteed if the corresponding first Chern class
$l^{-1}dA$ lies in the integer cohomology $H^2_{\rm de Rahm}(B_4,\bbZ)$.
It is straightforward to first check that $l^{-1}dA$ is indeed a globally
defined two-form on $B_4$. We next need to check that periods are
integral. A basis for the free part of the homology on $B_4$ is given by
$\Sigma_f$, the $(y,\psi)$ two-sphere fibre at a point on the round $S^2$, and $\Sigma_1$, $\Sigma_2$, the
two-spheres located at the poles $y=y_1$, $y=y_2$, respectively.
We note that we have the relation
$\Sigma_1=\Sigma_2-2\Sigma_f$ in homology.
If we denote the periods
for $\Sigma_f$ and $\Sigma_2$ to be integers $-q$ and $p$, respectively, we conclude that
must have
\bea\label{manu}
g(y_2)-g(y_1)&=&-lq\nn
g(y_2)&=&\frac{lp}{2}\ .
\eea
We note that the period for $\Sigma_1$ is then $p+2q$, consistent with the relation
between the two-cycles noted above. These conditions are satisfied if
\bea
\beta&=&\frac{q}{p+q}\nn
l&=&\frac{2(p+q)}{p(p+2q)}
\eea
with $p,q>0$. We choose $p$ and $q$ to be relatively prime
and then $Y_7$ is the product of $T^2$ with a simply connected manifold $M_5$.
By following the argument in \cite{Gauntlett:2004yd} we conclude that
topologically $M_5$ is $S^2\times S^3$.

Recalling that the circle bundle (parametrised by $z$)
is trivial over the two cycle $q\Sigma_2+p\Sigma_f$ we conclude that setting
$z$ to be constant, $q\Sigma_2+p\Sigma_f$ generates
$H_2(M_5,\bbZ)$. We also observe that $M_5$ has three obvious
three-cycles: $E_1$ and $E_2$ obtained by fixing $y=y_1$ or $y=y_2$, i.e. the
circle bundle over $\Sigma_1$ and $\Sigma_2$, and the three-cycle $E_3$
obtained by fixing a point on the round $S^2$, i.e. the circle bundle over
$\Sigma_f$. If we let $E$ be the generator of $H_3(M_5,\bbZ)$ we have
$E_1=-pE$, $E_2=-(p+2q)E$ and $E_3=-qE$. The generator $E$ can be obtained, for example,
as the linear combination $E=e_1E_1+e_2E_3$ where $e_1$ and $e_2$ are integers satisfying
$e_1p+e_2q=-1$.

At this stage we have shown that for each pair of relatively prime positive integers, $(p,q)$,
we have a regular manifold $Y_7=M_5\times T^2$ with $M_5=S^2\times S^3$.
In order to get a good solution of type IIB string theory
we now demand that the five-form flux is properly
quantised:
\bea N(D)=\frac{1}{(2\pi l_s)^4g_s}\int_DF_5\in
\bbZ \eea for any five-cycle $D\in H_5({Y}_7,\bbZ)$.
There are two independent five-cycles, $M_5$ at a fixed point on $T^2$ and
$S^3\times T^2$. For the latter, the $S^3$ factor is the generator $E$ of
$H_3(M_5,\bbZ)$, at a fixed point on the $T^2$. It is illuminating to calculate the
flux through the five-cycles $E_i\times T^2$, where the $E_i$ are the three-cycles on
$M_5$ introduced in the last paragraph.  After setting
\bea
\frac{L^4}{4\pi g_sl_s^4}&=&\frac{qp^2(p+2q)^2}{(p+q)^4}N\nn
Vol(T^2)&=&\pi\frac{q(p+q)^2}{p(p+2q)}\frac{M}{N} \eea
where $M$ and $N$ are integers,
we find that
\bea
\frac{1}{(2\pi l_s)^4g_s}\int_{M_5}F_5&=&-N\nn
\frac{1}{(2\pi l_s)^4g_s}\int_{E_1\times T^2}F_5&=&-pM\nn
\frac{1}{(2\pi l_s)^4g_s}\int_{E_2\times T^2}F_5&=&-(p+2q)M\nn
\frac{1}{(2\pi l_s)^4g_s}\int_{E_3\times T^2}F_5&=&-qM\ .
\eea
We see that the results are consistent with the relations in homology between
the three-cycles $E_i$ on $M_5$ that we noted above: in particular the five-form
flux through the cycle $E\times T^2$ is $M$.

We are now in a position to calculate the central charge of the
corresponding dual $d=2$ $(0,2)$ SCFT. Using \reef{bh} and \reef{bhconv}
we find that
\be\label{centralm}
c=6\frac{pq^2(p+2q)NM}{(p+q)^2}\ .
\ee

\subsection{Type IIB solutions with $G\ne 0$}

Let us now consider the solutions with $Q\ne 0$ and hence non-vanishing $G$.
The roots of $U$ are now given by
\be
y_{1,2}=\frac{1\mp \beta\sqrt{1+Q^2(\beta^2-1)}}{1+Q^2\beta^2}
\ee
and in order that we have two positive distinct roots, $y_2>y_1>0$ we demand that
\be
0<\beta^2< 1,\qquad
0\le Q^2< \frac{1}{1-\beta^2}\ .
\ee

We will again argue that $Y_7=M_5\times T^2$ with $M_5$ a circle fibration,
with the fibre coordinate labelled by $z$, over a four-dimensional base manifold, $B_4$,
parametrised by $y,\psi$ and the round $S^2$.
To ensure that $y,\psi$ parametrise a two-sphere, remarkably, it is again sufficient to choose $\psi$ to have period $2\pi$.
This again leads to a regular $B_4$, which is again topologically $S^2\times S^2$.
Following the logic of the last subsection, and calculating the periods of $l^{-1}dA/(2\pi)$,
to ensure that we have a good circle fibration over $B_4$
we now impose
\bea\label{manu2}
g(y_2)-g(y_1)&=&-lq\equiv -(lp)/X\nn
g(y_2)&=&\frac{lp}{2}
\eea
for relatively prime integers $p$ and $q$ and we have defined $X=p/q$.

Let us first consider $Q\ne 1$.
If $X>0$ we choose $Q<1$ and if $-1<X<0$ we choose $Q>1$ (other choices for $X$ lead to
the same solutions). We have
\bea
\beta^2&=&\frac{1-Q^2}{ (1+X)^2-Q^2}\nn
l&=&\frac{2((1+X)^2-Q^2)}{p(2+X)(1+X)}
\eea
and
\bea
y_1&=&\frac{X(1+X+Q^2)}{ (1+X)^2-Q^4}\nn
y_2&=&\frac{(2+X)(1+X-Q^2)}{ (1+X)^2-Q^4}\ .
\eea
Topologically $M_5=S^2\times S^3$. For future reference, we note that
as in the last subsection, the generator of $H_2(M\bbZ)$ is given by $q\Sigma_2+p\Sigma_f$ at fixed $z$.
Also as in the last subsection, $M_5$ has three natural three-cycles $E_i$ and the generator $E$ of $H_3(M_5,\bbZ)$,
is a linear combination of them.

For $Q=1$ we observe that
\bea
y_1=\frac{1-\beta^2}{1+\beta^2},\qquad y_2=1\ .
\eea
We further observe that $g(y_2)=0$ and hence we just need to demand that
the period of $l^{-1}dA/(2\pi)$ over $\Sigma_f$, the two sphere fibre parametrised by $y,\psi$,
is quantised which can be achieved by choosing
\be\label{manu3}
l=\frac{2}{1-\beta^2}\ .
\ee
For $Q=1$, the topology of $M_5$ is again $S^2\times S^3$, but the details are slightly different,
since the $z$ circle is only fibred over $\Sigma_f$.
For future reference, we can take $\Sigma_2$ to generate $H_2(M_5,\bbZ)$ and similarly, we can take the $z$ circle
fibred over $\Sigma_f$ to represent $H_3(M_5,\bbZ)$.

We have now shown that it is possible to switch
on the three-form flux and obtain infinite classes of regular
geometries. Furthermore, we observe that the five-form and the
three-form are globally defined on $Y_7$.

In order to find good solutions of string theory we need to ensure that
the three-form is suitably quantised. Writing $G=-dB-idC^{(2)}$ (since the axion and dilaton
are zero), we need to demand that
\bea
\frac{1}{(2\pi l_s)^2g_s}\int dC^{(2)}&\in& \bbZ\nn
\frac{1}{(2\pi l_s)^2}\int dB&\in& \bbZ\ .
\eea
Due to the Bianchi identity
\be
d{F_5}=\frac{i}{2}G\wedge G^*
\ee
we also need to ensure that corresponding Page charges (see e.g. \cite{Page:1984qv,Marolf:2000cb})
are quantised.
We will not carry out this analysis here, but an equivalent analysis will be carried out in
\cite{ajj} using the results of the next subsection.

\subsection{T-dual solutions}
After carrying out T-dualities along each of the two legs of the $T^2$, using the formulae in
appendix B, we arrive at the following type IIB solutions. The string frame metric is given by
\be
\frac{1}{\bar L^2}ds^2_\sigma=\frac{\beta}{y^{1/2}}[ds^2(AdS_3)+ds^2(X_7)]
\ee
where
\bea
ds^2(X_7)&=&\frac{\beta^2-1+2y-Q^2y^2}{4\beta^2}Dz^2+\frac{U(y)}{4(\beta^2-1+2y-Q^2y^2)}D\psi^2
+\frac{dy^2}{4\beta^2y^2 U(y)}\nn
&&+\frac{1}{\beta^2}ds^2(S^2)+ (du^1-\frac{Qy}{2\beta}[(1-g)D\psi-Dz])^2+(du^2)^2\ .
\eea
The dilaton is given by
\be
e^{2\phi}=\frac{\beta^2}{y}
\ee
and the RR three-form field strength is
\bea\label{threeformtdual}
\frac{1}{\bar L^2}dC^{(2)}&=&-\frac{1}{4\beta^2}dy\wedge D\psi\wedge Dz
-\frac{y}{\beta^2}J\wedge Dz+[\frac{1-yg}{\beta^2}]J\wedge D\psi\nn
&+&\frac{Q}{2\beta}du^1\wedge [dy\wedge Dz -4yJ-(1-g)dy\wedge D\psi]+2 Vol(AdS_3)\ .
\eea
Note that $\bar L$ is an arbitrary length scale that will be fixed by considering quantisation of the flux.

After a further $S$-duality transformation we obtain $AdS_3$ solutions with only NS fields non-vanishing,
but we will continue to work with the above solution.

For these solutions to be good solutions of
type IIB string theory
we need to ensure that the metric extends to a metric
on a globally defined manifold $X_7$ and that both the electric and magnetic
RR three-form charges are properly quantised:
\bea
n_1=\frac{1}{(2\pi l_s)^6 g_s}\int_{X_7} *dC^{(2)} \in \bbZ
\eea
and
\be
\frac{1}{(2\pi l_s)^2g_s}\int_T dC^{(2)}\in \bbZ
\ee
when integrated over any three-cycle $T\in H_3(X_7,\bbZ)$.

It is useful to note that since
\be
\frac{1}{\bar L^6}*dC^{(2)}=\frac{1}{4\beta^2 y^2}J\wedge dy\wedge D\psi\wedge Dz\wedge du^1 \wedge du^2
+Vol(AdS_3)\wedge(\dots)
\ee
we have
\be\label{noneex}
n_1=\left(\frac{\bar L}{l_s}\right)^6\frac{l}{g_s64 \pi^3\beta^2}\Delta u^1 \Delta u^2\frac{y_2-y_1}{y_1y_2}\ .
\ee
Thus, for any good solution of type IIB string theory, the central charge can then be written
\be\label{centy}
c=6n_1\left(\frac{\bar L}{l_s}\right)^2\frac{1}{g_s}\ .
\ee
To get the explicit expression we need the values of $\Delta u^1$, $\Delta u^2$ and $\bar L^2$.
In this paper we will only analyse this further for the case of $Q=0$, recovering results compatible with those
of the last subsection. The analysis for the case of $Q\ne 0$ will be carried out in \cite{ajj}.

\subsubsection{$Q=0$}
When $Q=0$, we first observe that $ds^2(X_7)$ is precisely
the same as $ds^2(Y_7)$ in \reef{mettues}. In section 4.1 we showed that
$X_7=M_5\times T^2$ where $M_5$ is a manifold parametrised by
$z,\psi,y$ and the round $S^2$ and the $T^2$ is parametrised by $u^1$ and
$u^2$. Further $M_5=S^2\times S^3$.

Let us now consider the quantisation of the three-form on $X_7$.
After fixing a point on the torus, the three-cycles $E_i$ on $M_5$, introduced in section 4.1,
all give rise to three cycles on $X_7$. If we choose the length scale to satisfy
\be\label{lengthy}
\frac{1}{g_s}\left(\frac{\bar L}{l_s}\right)^2=\frac{pq^2(p+2q)M}{(p+q)^2}
\ee
where $M$ is an integer then we calculate
\bea
\frac{1}{(2\pi l_s)^2g_s}\int_{E_1} dC^{(2)}&=&-pM\nn
\frac{1}{(2\pi l_s)^2g_s}\int_{E_2} dC^{(2)}&=&-(p+2q)M\nn
\frac{1}{(2\pi l_s)^2g_s}\int_{E_3} dC^{(2)}&=&-qM\ .
\eea
In particular we see that the flux through the generator of $H_3(X_7, \bbZ)$, the three-cycle $E$ introduced in section 4.1 at a fixed
point on the torus, is $M$.

The expression \reef{noneex} takes the more explicit form
\be
n_1=\left(\frac{\bar L}{l_s}\right)^6\frac{1}{g_s16 \pi^3}Vol(T^2)\frac{ (p+q)^4}{p^2q(p+2q)^2}
\ee
which, after substituting \reef{lengthy}, provides a quantisation condition on $Vol(T^2)$.
For the central charge, after substituting \reef{lengthy} into \reef{centy},
we now recover the previous result \reef{centralm} (with $N=n_1$), as expected.

The fluxes that we have activated, plus the amount of supersymmetry
preserved, suggests that the dual SCFT might arise by taking configurations
of fundamental strings intersecting NS fivebranes with the other four
directions of the NS fivebranes wrapped on a
holomorphic four-cycles inside a Calabi-Yau four-fold.

\subsubsection{$Q\ne 0$}
A careful analysis of the topology of $X_7$ and the quantisation of the three-form flux when $Q\ne 0$
will be carried out in \cite{ajj}.

\section{Fibration Constructions using $KE$ spaces: $D=11$ solutions}

In this section we will present new $AdS_2$ solutions of $D=11$ supergravity with magnetic four-form flux switched on.
We take the local eight-dimensional dimensional K\"ahler metric, $ds^2_8$, to be
the product of $T^2$ with a six-dimensional local K\"ahler metric which is constructed using the line bundle over
a four dimensional K\"ahler Einstein space with positive curvature. We have presented a few details of the derivation of these solutions in appendix D.

The metric of $D=11$ supergravity is given by
\be
\frac{1}{L^2}ds^2=\frac{1}{6^{4/3}\beta^{2/3}y^{4/3}}[ds^2(AdS_2)+ds^2(Y_9)]
\ee
where $L$ is an arbitrary length scale,
\bea
ds^2(Y_9)=&&(1-8\beta y+12\beta y^2-4\beta Q y^4)Dz^2+\frac{4\beta y U(y)}{(1-8\beta y+12\beta y^2-4\beta Q y^4)}D\psi^2\nn
&&+\frac{9\beta}{yU(y)}dy^2
+36\beta yds^2(KE_4^+)+36\beta y^2ds^2(T^2)
\eea
with $D\psi=d\psi+2V$, $dV=2J_{S^2}$  and the metric on the four-dimensional positively curved K\"ahler-Einstein space, $ds^2(KE_4^+)$,
is normalised so that ${\cal R}_{KE}=6J_{KE}$.
We also have
\be
Dz=dz-g(y)D\psi
\ee
with
\be
g(y)=-\frac{2\beta y(1-3y+2Qy^3)}{1-8\beta y+12\beta y^2-4 \beta Q y^3}
\ee
and
\bea
U(y)=1-9\beta y(1-y)^2-Qy^3
\eea
with $\beta, Q$ constants.

Writing the four-form as
\bea G_4=AdS_2\wedge
F_2+F_4 \eea
we have
 \bea
\frac{1}{L^3}F_2=-J_{KE}-\frac{2}{y^3}dy\wedge Dz+\frac{2g}{y^3}dy\wedge D\psi-\frac{i}{2}du\wedge d\bar u
\eea
and
\bea
\frac{1}{L^3}F_4&=&6\beta^{1/2}Q\Bigg(2J_{KE}\wedge J_{KE}+\frac{1}{3}[(1-g)D\psi-Dz]\wedge J_{KE}\wedge dy\nn
&&-2iy^2J_{KE}\wedge du\wedge d\bar u
-\frac{iy}{3}dy\wedge[(1-g)D\psi-Dz]\wedge du\wedge d\bar u\Bigg)\ .
\eea

We will not carry out a complete analysis of these solutions, but it is clear that there are
infinitely many new regular solutions. As in the last section, the task is to choose appropriate values of the constants
$\beta, Q$ and ranges of the coordinates so that $Y_9$ is a $U(1)$ fibration, with fibre parametrised by $z$,
over an eight dimensional base manifold, parametrised by $\psi, y$, the $KE_4^+$ space and the two-torus.
By choosing appropriate $\beta, Q$ we can restrict $y$ to lie between two suitable roots of the cubic $U=0$.
One can then show that if $\psi$ has period $2\pi$, then, remarkably, the eight-dimensional base manifold
is a regular $S^2$ bundle, with $S^2$ parametrised by $y,\psi$, over $KE_4^+\times T^2$.
Demanding that the $U(1)$ fibration is well defined, for appropriately chosen period for $z$, will lead to
additional restrictions on the parameters, but it is clear that there will be infinite number of solutions.
Finally, there will be additional restrictions imposed by demanding that the four-form flux Page charges
are suitably quantised.

We conclude this section by pointing out that when $F_4=0$, i.e. when $Q=0$, if we dimensionally reduce on one leg of the
$T^2$ and T-dualise on the other, we obtain type IIB $AdS_3$ solutions as constructed
in \cite{Gauntlett:2006af} (see appendix A and section 3.1 of \cite{Gauntlett:2006ns}).
However, when $F_4\ne 0$, while we still get type IIB solutions, because $F_4$ has a term proportional to the volume of the torus,
the metric will no longer be a warped product of $AdS_3$ with a seven manifold.

\section{Conclusions}

We have analysed new general classes of supersymmetric $AdS_3$ solutions of type IIB supergravity and
$AdS_2$ solutions of $D=11$ supergravity, which are dual to SCFTs with $(0,2)$ supersymmetry in $d=2$ and
supersymmetric quantum mechanics with two supercharges, respectively. The constructions which generalise those of
\cite{Kim:2005ez,Kim:2006qu} to allow for additional fluxes, depend crucially on the ``transgression terms'' appearing
in the Bianchi identities.

We also presented a rich set of new explicit examples using some constructions that generalise
those of \cite{Gauntlett:2006ns}. For the type IIB $AdS_3$ solutions we found an infinite class of solutions with
vanishing three-form flux in section 3.1 and determined the central charge of the dual SCFT.
In section 4 we presented a different class of explicit solutions of type IIB, with
the three-form flux labelled by $Q$.
The solutions have a two-torus and after two T-dualities and an S-duality  we showed that the solutions
can be written in terms of NS fields only.
For the case when $Q=0$ we showed that the solutions extend to well defined solutions of type IIB
string theory and we calculated the corresponding central charge.
The analysis for the case of $Q\ne 0$ will be carried out in \cite{ajj}.

We also constructed analogous $AdS_2$ solutions of $D=11$ supergravity. It would
worthwhile carefully analysing the conditions required on the local solutions to give rise to
properly quantised solutions of M-theory.

Despite the richness of the constructions we have presented, it is clear that they can be generalised still further.
For example, the $D=11$ solutions in section 5 are constructed using a four-dimensional K\"ahler-Einstein manifold.
For the special case when this is $S^2\times S^2$ there are almost certainly generalisations when we allow
the ratio of the curvatures of the two $S^2$'s to vary.

It remains an important outstanding problem to identify the dual SCFTs for all of these examples.
For the classes of type IIB $AdS_3$ solutions that depend on NS fields only, it would also be very interesting
to construct the worldsheet CFT describing the type IIB solutions.

We also showed how the general class of $AdS$ solutions can be analytically continued to obtain general classes of
1/8 BPS bubble solutions with additional fluxes to the classes of solutions considered in \cite{Gauntlett:2006ns}.
It would be interesting to study these further. For example, the constructions of this paper can be used
to obtain explicit solutions.

\subsection*{Acknowledgements}
We would like to thank David Ridout,
Volker Schomerus, Daniel Waldram and
especially James Sparks for helpful discussions.
AD and NK would also like to thank the Institute for Mathematical Sciences at
Imperial College and JPG would like to thank the
Perimeter Institute for hospitality. JPG is supported by an
EPSRC Senior Fellowship and a Royal Society Wolfson Award.
NK is supported by the Science Research Center Program of the
KOSEF through the Center for Quantum Spacetime (CQUeST) of Sogang
University with grant number R11-2005-021, and by the Korea Research Foundation Grant No. KRF-2007-331-C00072.

\appendix
\section{$AdS$ solutions}

\subsection{$AdS_3$ solutions of type IIB supergravity}

We will be interested in bosonic configurations of type IIB
supergravity with constant axion and dilaton. For simplicity we will
mostly set the axion and dilaton to zero. We will use the
conventions for type IIB supergravity that were used in
\cite{Gauntlett:2005ww}. The conditions for such a configuration to
be supersymmetric read:
\bea
\nabla_{M}\epsilon-\frac{1}{96}\left(\Gamma_{M}{}^{P_{1}P_{2}P_{3}}G_{P_{1}P_{3}P_{3}}-9\Gamma^{P_{1}P_{2}}G_{MP_{1}P_{2}}\right)\epsilon^{c}\qquad\qquad&&\nn
+\frac{i}{16\cdot 5!}\Gamma^{M_{1}\ldots M_{5}}F_{M_{1}\ldots
M_{5}}\Gamma_{M}\epsilon&=0,
\label{eq:10d_diff_kill}\\
\Gamma^{P_{1}P_{2}P_{3}}G_{P_{1}P_{2}P_{3}}\epsilon&=0\label{eq:10_alg_kill}
, \eea where $F_5$ is self-dual, $F_5=*_{10}F_5$ and the complex
three-form $G$ can be written\footnote{If one changes the sign of
$C^{(2)}$ one gets the conventions used in \cite{Myers:1999ps}
.}\bea G &=&ie^{\phi/2}\left(\tau dB-dC^{\left(2\right)}\right), \nn
\tau&=&C^{(0)}+ie^{-\phi} . \eea We have also chosen
$\Gamma_{11}\epsilon  =-\epsilon$ where $\Gamma_{11}
=\Gamma_{0}\ldots\Gamma_{9}$. and we take $\epsilon_{0\dots 9}=+1$.
To obtain a supersymmetric solution to the equations of motion it is
sufficient \cite{Gauntlett:2005ww} to also impose
\bea
\nabla^{P}G_{MNP} &
=&-\frac{i}{6}F_{MNP_{1}P_{2}P_{3}}G^{P_{1}P_{2}P_{3}}
\label{eq:three_form_eq_motion}\\
G_{P_{1}P_{2}P_{3}}G^{P_{1}P_{2}P_{3}}&=&0\label{eomtwo}\\ dG & =&0\\
dF & =&\frac{i}{2} G\wedge G^{\ast}  \label{eq:equation_of_motion6}
\eea and at most one component of the Einstein equations, which is
automatically solved for the classes of solutions we consider.

We now introduce the following ansatz
\bea
ds^{2} &
=& e^{2A}ds^{2}\left(AdS_{3}\right)+ds_{7}^{2} , \nn
F_{5} &
=&\left(1+\ast_{10}\right)\text{Vol}({AdS_{3}})\wedge F_2 ,
\eea
as in \cite{Kim:2005ez}, but generalised  to include a closed
three form $G$ defined on the seven dimensional space. We also
demand that the Killing spinors  are the same as those for the
$AdS_3$ solutions with $G=0$ that were analysed in
\cite{Kim:2005ez}.

For the gamma matrices we take
\bea \Gamma_{\mu} &
=&\sigma_{1}\otimes\boldsymbol{I}_{8\times8}\otimes\tau_{\mu},\quad\mu=0,1,2\nn
\Gamma_{a} &
=&\sigma_{2}\otimes\gamma_{a}\otimes\boldsymbol{I}_{2\times2},\quad
a=3,\ldots,9\eea
where $\sigma_i$ are Pauli matrices
and we choose the three-dimensional and seven dimensional gamma matrices $\tau_\mu$ and
$\gamma_a$, respectively, to satisfy
\bea \tau_{0}\tau_{1}\tau_{2} &
=-\boldsymbol{I}_{2\times2},\nn \prod_{a}\gamma_{a} &
=-i\boldsymbol{I}_{8\times8}.\eea
For the Killing spinor $\epsilon$
we make the ansatz
\be
\epsilon
=\chi\otimes\eta\otimes \psi_{n}^{\left(i\right)}\label{eq:spinor_ansatz}
\ee
where $\chi$ is a constant spinor satisfying
\be
\sigma_{3}\chi  =\chi
\ee
$\psi_{n}^{\left(i\right)}$ are Killing spinors on $AdS_3$ satisfying
\be
\hat{\nabla}_{\mu}\psi_{n}^{\left(i\right)}
=\frac{n}{2}\tau_{\mu}\psi_{n}^{\left(i\right)},\quad n=\pm1,\,
i=1,2,
\ee
and $\eta$ is a seven dimensional Dirac spinor. After substituting
into \eqref{eq:10d_diff_kill} we find
the following system of equations
\begin{align}
\nabla_{a}\eta-\frac{1}{16}{e^{-3A}}\not\hskip -3pt F_{2}\gamma_{a}\eta & =0\label{eq:diff_7d}\\
\left(\frac{n}{2}e^{-A}+\frac{i}{2}\not\hskip -2pt\partial A+\frac{i}{16}e^{-3A}\not \hskip-3ptF_{2}\right)\eta & =0\label{eq:alg_7d}\\
\gamma^{p_{2}p_{3}}G_{ap_{2}p_{3}}^{\ast}\eta & =0\label{eq:G2_7d}\\
\gamma^{p_{1}p_{2}p_{3}}G_{p_{1}p_{2}p_{3}}\eta & =0.\label{eq:G3_7d}
\end{align}

As shown in \cite{Kim:2005ez}, by just using equations
\eqref{eq:diff_7d} and \eqref{eq:alg_7d}, the geometry and five form
flux are constrained to take the local form
\bea ds^{2}&= &
e^{2A}\left[ds^{2}\left(AdS_{3}\right)+\frac{1}{4}\left(dz+P\right)^{2}+e^{-4A}ds^2_6\right]\nn
F_{2}&= & {2n}J-\frac{1}{2}d\left[e^{4A}\left(dz+P\right]\right)
\label{10dmetric} \eea where $\partial_z$ is a Killing vector,
$ds^2_6$ is a six dimensional K\"ahler metric with K\"ahler form
$J$, Ricci form given by $\mathcal{R}=n\, dP$, scalar curvature
$R=8e^{-4A}$ and holomorphic three form $\Omega$. This result is
obtained by analysing various bilinears in $\eta$. In particular we
note that
$\eta^\dagger\eta=e^A$, $\Omega=e^{2A}e^{i nz}\eta^T\gamma_{(3)}\eta$ and
$J=-e^{A}i\eta^\dagger\gamma_{(2)}\eta$. Furthermore, $K\equiv
\eta^\dagger \gamma_{(1)}\eta=(e^{2A}/2)(dz+P)$, so that the
corresponding dual vector is the Killing vector $2\partial_z$. It
is also useful to note that $\Psi\equiv\eta^T\gamma_{(4)}\eta=-
e^{-3A}e^{-ni z}K\wedge \Omega$.

We next argue that \be i_{K}G  =0 .  \label{eq:alongK} \ee
To see this we first multiply \reef{eq:G2_7d} by
$\eta^T\gamma_k\gamma_a$ and \reef{eq:G3_7d} by $\eta^T\gamma_k$ to
deduce that \be \Omega_k{}^{p_1p_2}(i_KG)_{p_1p_2}=0,\qquad
\bar\Omega_k{}^{p_1p_2}(i_KG)_{p_1p_2}=0\ . \ee This shows that the
$(0,2)$ and $(0,2)$ pieces of $i_KG$ vanish. Next multiplying
\reef{eq:G2_7d} by $\eta^T\gamma_{q_1q_2q_3}$ we deduce that \be
\bar\Omega^p{}_{[q_1 q_2 }G_{q_3]pr}=0\ . \ee Letting $q_1$ be just in
the $z$ direction we deduce that
\be \bar\Omega^p{}_{q_1 q_2
}(i_KG)_{pr}=0 \ee showing that the $(1,1)$ piece of $i_KG$ also
vanishes.

Since $i_KG=0$ we can now decompose $G$ in terms of
$\left(p,q\right)$ forms on $\mathcal{B}_{6}$\[
G=G^{\left(1,2\right)}+G^{\left(2,1\right)}+G^{\left(3,0\right)}+G^{\left(0,3\right)}.\]
From equations \eqref{eq:G2_7d} and \eqref{eq:G3_7d} we obtain\be
\Omega^{p_{1}p_{2}p_{3}}G_{p_{1}p_{2}p_{3}}  =0, \qquad
\bar{\Omega}_{\qquad a}^{p_{1}p_{2}}G_{p_{1}p_{2}b}  =0, \ee
implying that only the $\left(1,2\right)$ component of the three
form $G$ can be non-zero. From equation \eqref{eq:G2_7d} we have
that \be J\wedge G  =0 . \ee Thus we conclude that supersymmetry
implies that the $\left(1,2\right)$ form $G$ is primitive. These two
properties when combined give the duality condition on the base
$\mathcal{B}_{6}$
\begin{equation}
\ast_{6}G=i G\label{eq:three_form_dual},
\end{equation}
where we used the volume form \be \mathrm{Vol}_{6}=\frac{1}{6}\,
J\wedge J\wedge J. \ee We can now easily check that
\eqref{eq:three_form_eq_motion} and \reef{eomtwo} are both
satisfied.

Thus to ensure that all equations of motion are satisfied we just
need to ensure that \eqref{eq:equation_of_motion6} holds.
Using \eqref{eq:three_form_dual} we find that
\eqref{eq:equation_of_motion6} can be written as\begin{equation}
\frac{1}{16}J\wedge\mathcal{R}\wedge\mathcal{R}+\frac{1}{32}d\ast_{6}dR=-\frac{1}{8}G\wedge\ast_{6}G^{\ast},\end{equation}
which may also be written as a scalar equation\begin{equation} \Box
R-\frac{1}{2}R^{2}+\mathcal{R}^{ij}\mathcal{R}_{ij}+\frac{2}{3}G^{ijk}G_{ijk}^{\ast}=0.\label{eq:equation_motion6_scalar}\end{equation}

Note that in the main text we have fixed $n$ to be +1. The solution preserves four supersymmetries
since $i$ runs from 1 to 2 in the $AdS_3$ Killing spinors $\psi^{(i)}_n$ appearing in \reef{eq:spinor_ansatz} and $\eta$ is a Dirac spinor.
Two of these are Poincar\'e supersymmetries and two are special conformal supersymmetries.
Using horospherical coordinates, the Poincar\'e Killing spinors on $AdS_3$ are eigenvalues of the gamma matrix along the radial direction, say $\tau_2$
\cite{Lu:1996rhb}. Observing that $\Gamma_{01}=-\boldsymbol{I}_{2\times 2}\otimes \boldsymbol{I}_{8\times 8}\otimes \tau_2$ we see that the two Poincar\'e
supersymmetries are eigenvalues of $\Gamma_{01}$ with the same eigenvalue and hence the solutions are dual to SCFTs with $(0,2)$ supersymmetry.

\subsection{$AdS_2$ solutions of $D=11$ supergravity}

The condition for a bosonic configuration of $D=11$ supergravity to
be supersymmetric reads \be \delta\psi_{M}
=\nabla_{M}\epsilon+\frac{1}{288}\left[\Gamma_{M}{}^{N_{1}N_{2}N_{3}N_{4}}-8\delta_{M}^{N_{1}}\Gamma^{N_{2}N_{3}N_{4}}\right]G_{4N_{1}N_{2}N_{3}N_{4}}\epsilon=0\label{eq:11d_Killing}
, \ee where we are using the conventions of \cite{Gauntlett:2002fz}
and in particular $\Gamma_{0\ldots10} =1$ and $\epsilon_{0\dots
10}=+1$. For the supersymmetric bosonic configurations we will be
considering, in order that all equations of motion are satisfied it
is sufficient \cite{Gauntlett:2002fz} to also just demand that \bea
dG_4&=&0,\nn d\ast_{11}G_4 & =&-\frac{1}{2}G_4\wedge
G_4\label{eq:11d_eq_motion}. \eea

Our $AdS_{2}$ ansatz is
\begin{align}
ds^{2} & =e^{2A}ds^{2}\left(AdS_{2}\right)+ds_{9}^{2},\nn G_{4} &
=\mathrm{Vol}({AdS_{2}})\wedge F_{2}+{F}_{4},
\end{align}
where $F_2$ and ${F}_{4}$ are closed forms defined on the nine
dimensional space. For the gamma matrices we perform the
reduction\begin{align} \Gamma_{\mu} &
=\tau_{\mu}\otimes\boldsymbol{I},\quad\mu=0,1\nn \Gamma_{a} &
=\tau_{2}\otimes\gamma_{a},\quad a=2,\ldots,10\end{align} with
$\tau$ and $\gamma$ being real matrices and we use the conventions
\begin{align}
\tau_{0}\tau_{1}\tau_{2} & =-1, \nn \prod_{a}\gamma_{a} &
=-1.\end{align} In this representation we can make the ansatz for
the eleven dimensional Majorana spinor
\be
\epsilon=\chi^{(i)}_{n}\otimes\eta+\mathrm{c.c.}
\ee
where the $\eta$ is a nine-dimensional Dirac spinors and the real three-dimensional
spinor $\chi^{(i)}_{n}$ satisfies\begin{equation}
\hat{\nabla}_{\mu}\chi^{(i)}_{n}=\frac{in}{2}\tau_{\mu}\tau_{2}\chi^{(i)}_{n},\quad i=\pm1,\,n=\pm1\label{eq:AdS2_Kill_spinor},
\end{equation}
and can be taken to satisfy the orthogonality condition
\begin{equation}
(\chi^{(i)}_{n})^{\dagger}\tau_{2}\chi^{(i)}_{n}=0\label{eq:Kill_spinor_orthogonality}.\end{equation}
(which can be checked, for example, by explicitly calculating the
spinors).

We now find the following system of equations
\begin{align}
\left[\nabla_{a}+\frac{1}{24}e^{-2A}\left(\gamma_{a}^{\;\; bc}F_{2bc}-4F_{2ab}\gamma^{b}\right)\right]\eta & =0\label{eq:diff_9d_Kill}\\
\left[in e^{-A}+\gamma^a\partial_a {A}-\frac{1}{6}e^{-2A}\gamma^{ab} F_{2ab}\right]\eta & =0\label{eq:alg_9d_Kill}\\
\gamma^{b_{1}b_{2}b_{3}}{F}_{4ab_{1}b_{2}b_{3}}\eta & =0.\label{eq:four_form_alg}\end{align}
Using the results of \cite{Kim:2006qu} one can show that equations \eqref{eq:diff_9d_Kill}
and \eqref{eq:alg_9d_Kill} imply that the metric and the two form flux are
constrained to be of the form
\begin{align}
ds^{2} & =e^{2A}\left[ds^{2}\left(AdS_{2}\right)+\left(dz+P\right)^{2}+e^{-3A}ds_{8}^{2}\right],\\
F_{2} & =n J+d\left[e^{3A}\left(dz +P\right)\right],\end{align} where $\mathcal{R}=-n dP$ and
$ds^2_8$ is K\"ahler with K\"ahler form $J$, Ricci potential given
by $P$ and scalar curvature given by $R=2e^{-3A}$.

The constraint \eqref{eq:four_form_alg} implies that the only non-zero part of
the magnetic component ${F}_{4}$ is a $\left(2,2\right)$ and primitive
form with no non-zero components along the $z$ direction:
\begin{align}
J\wedge{F}_{4} & =0,\\
i_{K}{F}_{4} & =0.
\end{align}
Here $K$ is the one-form constructed out of the nine dimensional
bilinears $K=\eta^{\dagger}\gamma_{(1)}\eta=e^{2A}\left(dz+P\right)$
whose dual is the Killing vector $\partial_z$. Note that these
conditions imply that the four form is also self-dual with respect
to $ds^2_8$: \be \ast_{8}{F}_{4}={F}_{4} . \ee Using that the $D=11$
epsilon tensor is given by
$\epsilon=-e^{-A}Vol(AdS_2)(dz+P)\frac{J^4}{4!}$,
we find that the equation of motion for the four form \reef{eq:11d_eq_motion} implies that
\begin{equation}
J^{2}\wedge\mathcal{R}\wedge\mathcal{R}+d\ast_{8}dR={F}_{4}\wedge{F}_{4}.\label{eq:equation_of_motion}\end{equation}
which may also be written as a scalar equation\begin{equation}
\Box R-\frac{1}{2}R^{2}+\mathcal{R}^{ij}\mathcal{R}_{ij}+\frac{1}{4!}F_{ijkl}F^{ijk}=0.\end{equation}

\section{T-duality}
We consider a type IIB solution with a square two-torus,
parametrised by $u^1$ and $u^2$, of the form \bea
ds^2&=&e^{2A}\left[ds^2(AdS_3)+ds^2(M_5)+\Sigma((du^1)^2+(du^2)^2)\right]\nn
F_5&=&f_5+f_3\wedge du^1\wedge du^2\nn G&=&(du^1-idu^2)\wedge dv\nn
\phi&=&0,\qquad C^{(0)}=0\eea where $f_5$, $f_3$, $v$ and
$ds^2(M_5)$ have no dependence on the coordinates $u^i$.
Using the formulae in, for example, \cite{Hassan:1999bv} we can
T-dualise on the $u^1$ direction and then the $u^2$ direction to get
the following type IIB solution \bea
ds^2_\sigma&=&e^{2A}\left[ds^2(AdS_3)+ds^2(M_5)\right]+\frac{1}{\Sigma
e^{2A}}\left[(du^1-v)^2+(du^2)^2\right]\nn dC^{(2)}&=&f_3-dv\wedge
(du^1-v)\nn e^{2\phi}&=&\frac{1}{\Sigma^2 e^{4A}}\eea where the
metric, here, is written in the string frame.

\section{Type IIB solutions from fibrations over $S^2\times T^2$}

Consider the following ansatz for a six dimensional K\"ahler metric
\be
ds_{6}^{2}=\frac{dx^{2}}{4x^3U(x)}+\frac{U(x)}{x} D\phi^{2}+\frac{1}{x}ds^{2}\left(S^2\right)+du\, d\bar u ,
\ee
where $D\phi=d\phi+V$, $dV=2J_{S^2}$, the $S^2$ is normalised so that
${\cal R}_{S^2}=4J_{S^2}$ and we have introduced a complex coordinate $u=u^1+iu^2$
for a $T^{2}$ factor.
In this case the K\"ahler form $J$ and the $(3,0)$ form $\Omega$ read
\bea
J&=&-\frac{1}{2x^2}d x\wedge D\phi +\frac{1}{x}J_{S^2}+\frac{i}{2}\, du\wedge d\bar u ,
\nn
\Omega&=&e^{2i\phi}\left[-\frac{1}{2x^2\sqrt U}dx+i\frac{\sqrt U}{x} D\phi\right]\wedge \Omega_{S^2}\wedge du .
\eea
We have $d\Omega=iP\wedge \Omega$ where $P$ is the Ricci form given by
\be
P=f D\phi,\qquad f=2(1-U)+xU' .
\ee
It is easy to calculate the Ricci form, given by ${\cal R}=dP$,  and we record that the
Ricci scalar is given by
\be
R  =4xf-4x^2f' .
\ee
For the three form $G$ we make the simple ansatz that it is the wedge product of
$d\bar u$ with a primitive $(1,1)$ form on the four-dimensional K\"ahler space parametrised by
$x,\phi$ and the $S^2$. This leads us to consider
\be
G=d\bar u\wedge d\left[qx D\phi\right]\ .
\ee

If we now substitute into \reef{miib}, after integrating once,
we are led to the following differential equation for $U$:
\be
2f^{2}+U\, R^{\prime}+8q^{2}x^{2}=constant .
\ee
We look for polynomial solutions to this equation by considering
the ansatz
$U (x) =1+\sum_{i=0}^{2}a_{i}x^{i}$. This implies that
$R  =-8a_{0}x$ and in order to have $R>0$ we choose $a_{0}=-1/\beta^2$.
A little calculation shows that
%
$U$ takes the form
\be
U(x)  =1-\frac{1}{\beta^2}\left(1-\frac{a_{1}\beta^2}{2}x\right)^{2}- q^{2}\beta^2 x^{2} .
\ee

It is now straightforward to assemble the full ten-dimensional solution using \reef{iiboa}-\reef{iib4}.
It is convenient to make the following rescalings
\be
y=\frac{a_1\beta^2}{2}x,\qquad Q=\frac{2}{a_1\beta}q, \qquad \tilde u=\frac{\sqrt 2}{\beta\sqrt a_1}u\ .
\ee
Furthermore we also perform a simultaneous scaling of the ten-dimensional metric and the three-form
by a factor of $\frac{\sqrt 2}{\beta\sqrt a_1}$ and the five-form by a factor of
$\frac{2}{\beta^2 a_1}$ (which indeed transforms a solution to another solution).
Finally, it is very helpful to perform the coordinate change $\phi=(\psi-z)/2$ and this then leads to the type IIB solutions
as recorded in the main text, although we note that we have dropped the tildes form the coordinates on the torus for clarity.

\section{$D=11$ solutions from fibrations over $KE_4^+\times T^2$}

Consider the following ansatz for an eight dimensional K\"ahler metric
\be
ds_{8}^{2}=\frac{dx^{2}}{4x^3U(x)}+\frac{U(x)}{x} D\phi^{2}+\frac{1}{x}ds^{2}\left(KE^+_4\right)+du\, d\bar u ,
\ee
where $D\phi=d\phi+V$, $dV=2J_{KE}$, the K\"ahler-Einstein four metric with positive curvature, $ds^2(KE^+_4)$,
is normalised so that ${\cal R}_{S^2}=6J_{S^2}$ and $u=u^1+iu^2$ is a complex coordinate
for a $T^{2}$ factor.
In this case the K\"ahler form $J$ and the $(4,0)$ form $\Omega$ read
\bea
J&=&-\frac{1}{2x^2}d x\wedge D\phi +\frac{1}{x}J_{KE}+\frac{i}{2}\, du\wedge d\bar u ,
\nn
\Omega&=&e^{3i\phi}\left[-\frac{1}{2x^{5/2}\sqrt U}dx+i\frac{\sqrt U}{x^{3/2}} D\phi\right]\wedge \Omega_{KE}\wedge du .
\eea
We have $d\Omega=iP\wedge \Omega$ where $P$ is the Ricci form given by
\be
P=fD\phi,\qquad f=3(1-U)+xU' .
\ee
It is easy to calculate the Ricci form, given by ${\cal R}=dP$,  and we record that the
Ricci scalar is given by
\be
R  =8xf-4x^2f' .
\ee
For the magnetic four form, $F_4$, we choose the ansatz:
\bea
F_4&=&A_2\wedge(J_6-\frac{i}{2}du\wedge d \bar u)\nn
&=&A_2\wedge(-\frac{1}{2x^2}dx\wedge D\phi+\frac{1}{x}J_{KE}-\frac{i}{2}du\wedge d \bar u)
\eea
where $J_6$ is the K\"ahler form on the six space excluding the torus.
We clearly have that $F_4$ is $(2,2)$ and is closed provided that the two-form $A_2$ is $(1,1)$ and closed.
A suitable ansatz is $A_2=d[\Phi(x)D\phi]$ and we find that $F_4$ is primitive provided that
$\Phi=qx^2$ for an arbitrary constant $q$. We thus have
\be
F_4=d[qx^2D\phi]\wedge(-\frac{1}{2x^2}dx\wedge D\phi+\frac{1}{x}J_{KE}-\frac{i}{2}du\wedge d \bar u)\ .
\ee

If we now substitute into \reef{md11}, after integrating once,
we are led to the following differential equation for $U$:
\be
4f^{2}+U\, R^{\prime}+4q^{2}x^{4}=constant\times x .
\ee
We look for polynomial solutions to this equation by considering
the ansatz
$U(x) =\sum_{i=0}^{3}a_{i}x^{i}$. We find two classes of solutions, one with $a_0=1$ and the other with $a_0=3$.
Since we are interested here in $AdS_2$ solutions, we only consider the solution with $a_0=1$ and we have
\be
U(x)=1+a_1x\left(1+\frac{a_2}{2a_1}x\right)^2+\frac{q^2}{4a_1}x^3\ .
\ee
Since $R=-8a_1x^2$, we demand that $a_1<0$.

It is now straightforward to assemble the full eleven-dimensional solution using \reef{pop}-\reef{pop4}.
It is convenient to make the following rescalings
\be y=\frac{-a_2}{2a_1}x,\qquad Q=\frac{2a_1^2}{a_2^3}q^2,\qquad \tilde u=\frac{\sqrt {-2a_1}}{\sqrt {a_2}}u\ .
\ee
We also define $\beta=\frac{2a_1^2}{9a_2}$.
Furthermore we also perform a simultaneous scaling of the eleven-dimensional metric
by a factor of $\left(\frac{2(-a_1)}{a_2}\right)^{2/3}$ and the four-form by a factor of
$\frac{2(-a_1)}{a_2}$ (which indeed transforms a solution to another solution).
Finally, it is very helpful to perform the coordinate change $\phi=(\psi-z)/3$ and this then leads to the $D=11$ solutions
as recorded in the main text, although we note that we have dropped the tildes form the coordinates on the torus for clarity.

\end{document}